\documentclass[11pt]{article} 

\usepackage{geometry} 
\usepackage[backend=bibtex, style=numeric]{biblatex}
\usepackage{graphicx} 
\usepackage{tikz}
\usepackage{float} 
\usepackage{wrapfig} 
\usetikzlibrary{arrows}
\usepackage{algorithm}
\usepackage[noend]{algorithmic}
\usepackage{amsmath,amssymb,amsthm}

\usepackage{ifsym}
\begin{document}


\begin{titlepage}

\newcommand{\HRule}{\rule{\linewidth}{0.5mm}} 

\center 

\textsc{\LARGE Oberlin College}\\[1.5cm] 
\textsc{\Large Computer Science Department}\\[0.5cm] 
\textsc{\large Honors Thesis}\\[0.5cm] 

\HRule \\[0.4cm]
{ \huge \bfseries Selfish Routing on Dynamic Flows}\\[0.4cm] 
\HRule \\[1.5cm]

\begin{minipage}{0.4\textwidth}
\begin{flushleft} \large
\emph{Author:}\\
Christine \textsc{Antonsen} 
\end{flushleft}
\end{minipage}
~
\begin{minipage}{0.4\textwidth}
\begin{flushright} \large
\emph{Supervisor:} \\
Alexa \textsc{Sharp} 
\end{flushright}
\end{minipage}\\[4cm]

{\large April 2, 2015}\\[3cm] 

\vfill 

\end{titlepage}



\small{\begin{abstract}
Selfish routing on dynamic flows over time is used to model scenarios that vary with time in which individual agents act in their best interest. In this paper we provide a survey of a particular dynamic model, the deterministic queuing model, and discuss how the model can be adjusted and applied to different real-life scenarios. We then examine how these adjustments affect the computability, optimality, and existence of selfish routings. 
\end{abstract}}
\tableofcontents
\newpage
\section{Introduction}
Dynamic flow networks can model traffic, optical networks, building evacuations, and more. In many of these situations individual agents using a network, road, or trying to escape a building want to maximize their personal welfare, and thus act selfishly. In the field of algorithmic game theory we measure the loss in quality of a solution where everyone acts selfishly compared to some optimal solution to a given objective function. For example, how bad is it when everyone chooses their own evacuation route from a building compared to a fire marshal directing everyone out of the building? Already there is a great deal of research on selfish routing on flows, but these flows have been static, meaning the model does not take into account time as a variable and flow traverses edges instantaneously. In dynamic flows it takes time for flow to traverse an edge, and the amount of flow, amongst other parameters, can vary with time. Only recently have people started to study selfish routing on these dynamic flow models. In this paper we focus on selfish routing in the deterministic queuing model and how minor changes in the model can affect the computability, optimality, and existence of selfish routings. \par
The current literature surrounding selfish routing on dynamic flows is growing, but is hard to navigate. It is very difficult to compare results from different papers which do not use consistent terminology, keep track of the various modifications and results within a single paper, or even abstract the results to broader instances. This paper serves to unify the current results on the deterministic queuing model, deepen the applications of the model, apply the results to broader categories, and provide some intuition along the way. The rest of this section presents the selfish routing model in the context without time, and provides definitions for terms that are needed in the rest of the paper. Section 2 presents the deterministic queuing model and the tools used to analyze the model, and then discusses several applications of the general model. Section 3 considers ways to modify the deterministic queuing model to better fit more applications. Finally, Section 4 appertains to the results in optimality, computability, and existence of the deterministic queuing model and its modifications. 
\subsection{Static Selfish Routing}
An instance of a selfish routing game is given by a directed graph $G=(V,E)$ and a set of source-sink vertex pairs, $(s_1,t_1),\ldots,(s_k,t_k)$ called commodities. Each player $i$ is associated with    
\begin{wrapfigure}{l}{0.375\textwidth} 
	\begin{center}
    		\begin{tikzpicture}[->,>=stealth',shorten >=.50pt,auto,node distance=3cm,thick,main node/.style={circle,draw}]
  			\node[main node] (s) {s};
  			\node[main node] (t) [right of=s] {t};					
			\path (s) edge  [bend right] node [below] {$c(x)=1$} (t)
					edge  [bend left] node {$c(x)=x$} (t);
		\end{tikzpicture}
	 \end{center}
 \caption{\small{Atomic Pigou's Example}}
\end{wrapfigure}
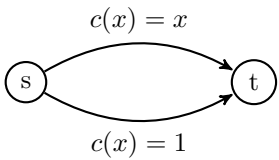
one commodity, and they have a specific amount of traffic $r_i$, also called flow,  that they need to route from $s_i$ to $t_i$. In addition each edge $e \in E$ of the graph has a nonnegative, continuous, and non-decreasing cost function $c_e(x)$, which represents the cost of traversing the edge when there is $x$ amount of traffic on it \cite{AGTRoutingGames}. Each player wants to minimize their total cost of routing, and we choose an objective function that minimizes the total cost incurred by all players.
In Figure 1, we can imagine four players each wanting to route $.25$ units of flow from the source $s$, to the sink $t$. Each player can either take the lower edge and incur a cost of one, as if they are on a superhighway whose traversal time is independent of the amount of traffic, or they can take the upper edge and incur a cost that equals the total amount of flow being routed on that edge. Since each player wants to minimize their own cost, all the players will use the upper edge: if any one player is on the lower edge, then the upper edge has total flow less than one, and thus is cheaper. In this situation the total cost incurred by all of the players is $4$ but routing two players on the top edge and the other two players on the bottom edge minimizes the total cost of all the players, a total cost of $3$. In the field of algorithmic game theory we like to compare how bad a situation is when everyone acts selfishly compared to some optimal solution to the problem.
\subsection{Algorithmic Game Theory Basics}
There are two different types of routing games. In \textbf{nonatomic} routing, there are a large number of players, each controlling a negligible amount of the total overall flow. In \textbf{atomic} routing, each player controls a significant amount of the overall flow. Atomic instances can either be weighted or unweighted; in an \textbf{unweighted} atomic instance every player controls the same amount of flow, whereas in \textbf{weighted} instances the amount of flow each player controls differs.\par
A \textbf{Nash equilibrium} is when given every other player's strategy, no player would be better off switching their current strategy. In the example given in Figure 1, a Nash equilibrium is all of the players routing their flow on the top edge. To measure the inefficiency of selfish routing we use the \textbf{price of anarchy}, defined as the ratio of the cost of the worst Nash equilibrium to the cost of an optimal routing. Intuitively, the price of anarchy quantifies the loss in quality when players act selfishly instead of being forced to behave optimally. The price of anarchy for the example in Figure 1 is $\frac{4}{3}$ because, as we explained in Section 1.1, the total cost of all the players in the Nash equilibrium is $4$ but the minimal cost routing had a total cost of $3$.
\section{The Deterministic Queuing Model} 
While selfish routing on static flows is a popular research area in the algorithmic game theory community, the model is unrealistic. Static flows do not take into account time as a variable and the flow traverses edges instantaneously; this is where dynamic flows come in. There are many different ways to model routing on dynamic flows, and all models incorporate time and parameter variance over time. We specifically focus on the deterministic queueing model because it can be modified to fit many different applications. This section presents the model, the objective functions and prices of anarchy that are used to analyze the model, and several applications of the model.
\subsection{The Model}
Vickrey introduced the deterministic queuing model \cite{vickrey}, which Koch and Skutella later made popular in the field of algorithmic game theory \cite{koch2011nash}. The model consists of a directed graph $G=(V,E)$, a single \textbf{source node} $s$, and a single \textbf{sink node} $t$. Each edge has a non-negative, Lebesgue integrable \textbf{capacity function} $u_e: \mathbb{R}_+ \rightarrow \mathbb{R}_+$, where $u_e(\theta)$ bounds the rate at which flow is able to leave edge $e$ at time $\theta$. If more flow particles want to leave an edge than its capacity allows then they form a waiting queue, which has no physical dimension. 
\begin{wrapfigure}{l}{0.475\textwidth} 
  \begin{center}
    \includegraphics[width=0.475\textwidth]{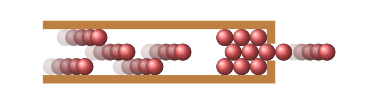}
  \end{center}
  \caption{Waiting Queue on an Edge.\newline \small{\fullcite{kochthesis}}}
\end{wrapfigure}
Figure 2 serves as a visual representation of the formation of a waiting queue. Each edge also has a constant \textbf{free flow transit time} $\tau_e \in \mathbb{R}_+$ which represents the time it takes to traverse edge $e$ if the waiting queue on $e$ is empty. Players arrive at the source $s$ at a fixed rate $d$, and as soon as a flow particle arrives at the source it determines what $s-t$ path to take and immediately begins to route its flow on that path. Since each player wants to arrive at their destination $t$ as quickly as possible, every flow particle tries to arrive at $t$ before the particles in front of it, while not being overtaken by the flow particles from behind. This results in no flow particle being overtaken, and thus the deterministic queuing model naturally obeys the first-in-first-out (FIFO) property. This also implies that flow particles never wait longer than they have to, meaning that they will always exit an edge if the capacity allows. 
\subsection{Objective Functions}
Dynamic flows have parameters that can vary with time and also include time as a parameter, allowing for many different types of objective functions. This paper focuses on three particular objective functions. Given an amount of flow $M$, a \textbf{quickest flow} sends that amount of flow from the source to the sink in the minimal amount of time, $T$. An \textbf{earliest arrival flow} maximizes the amount of flow that arrives at the sink for every interval $[0,t]$, where $0 \leq t \leq P$, where $P$ is some specified end time. An earliest arrival flow is also a quickest flow \cite{Stackelberg, kotnyek}. A player's \emph{bottleneck value} is the highest cost/latency incurred from an edge on their routing path. The \emph{bottleneck} of a flow routing is the highest bottleneck value of the players. A \textbf{narrowest flow} is a flow with minimum bottleneck value \cite{AtomicDQM}.
\subsection{Prices of Anarchy}
In \cite{kochthesis, Stackelberg}, the authors describe three common prices of anarchy that are used to study selfish routing. The \textbf{evacuation price of anarchy} compares the total amount of flow that has reached the sink by some time $\theta$ in the worst equilibrium, to an earliest arrival flow. We might study the evacuation price of anarchy if we were looking at building evacuation plans. The goal is to have as many people exit the building at every point in time $\theta$ because we do not always know how long we have to evacuate everyone. The evacuation price of anarchy will show us how bad it would be if everyone chose their own evacuation plan instead of listening to a central authority. The \textbf{total delay price of anarchy} compares the total delay of the worst equilibrium to an earliest arrival flow, where the total delay is the sum of all the players' arrival times at the sink. Given an amount of flow $M$, the \textbf{completion time price of anarchy} compares the amount of time it takes to route $M$ amount of flow from the source to the sink in the worst equilibrium to a quickest flow. \par
In \cite{AtomicDQM}, the authors seek to minimize the bottleneck of the game. For that reason, we introduce the \textbf{bottleneck price of anarchy} which compares the bottleneck value in the worst equilibrium to the narrowest flow. When we state that a price of anarchy is bounded above by some value $x$, that is saying that the price of anarchy will always be less than or equal to $x$. This also means that the value of the worst equilibrium will never be more than $x$ times the optimal value.
\subsection{Applications}
\subsubsection{Building Evacuations}
The deterministic queuing model can be used to model building evacuations. In an emergency we want to get as many people out of the building as fast as possible. Each individual exiting the building wants to get out as fast as possible, and thus acts selfishly. We can imagine hallways and stairways as the edges, and the time it takes to walk down a hallway, or down a flight of stairs can be depicted through the free flow transit time. Similarly, every single person cannot try to go through a doorway at the same time to exit a hallway, so people will crowd around the doorway, i.e., form a waiting queue, until there is room for them to get through. The capacities of the edges can thus depict the doorways.
\subsubsection{Traffic Forecasting}
The traffic community uses forecasting models to evaluate how changes in transportation facilities, demographics, and more impact the transportation system of a region. Before the introduction of Dynamic Traffic Assignment models, transportation planning used static models. Dynamic models are more realistic because they incorporate time into the model, and are thus useful in evaluating travel times and costs to both players and the entire system \cite{DTA}. An important factor of the model is simulating different types of congestion. One type of congestion is a bottleneck, in which one road segment leads into another road segment with a smaller capacity \cite{vickrey}. This type of congestion can be represented through the deterministic queuing model because the edge capacities limit how many players can exit that edge, and thus replicate a smaller amount of players being allowed onto the following edge. The use of the deterministic queuing model in this application would be better if we made the waiting queues have physical dimension. This way, when the waiting queue reaches its physical dimension it causes longer waiting on the incoming edges, which is what happens to roads that feed into traffic jams. Similarly, it would be interesting to modify the model so that there were multiple edges between two vertices representing different lanes of traffic since not all lanes of traffic move at the same speed, and because some lanes have better priority, such as an HOV lane.
\section{Modifications of the Deterministic Queuing Model}
An advantage of using the deterministic queueing model is that it is easily adapted to different real-life applications. This section examines different model modifications and the applications they pertain to. 
\subsection{Shortest Paths Networks}
A simpler version of the deterministic queuing model is called a shortest paths network and only involves slight modifications: each edge in the network has constant capacity, and each path from the source to the sink has the same total free flow transit time, i.e., the sum of the free flow transit times along the edges in a path from the source to the sink is the same no matter what path you use. Instead of flow arriving at the source at a fixed rate $d$, all flow units are present at the source at time $0$. In the example below we consider an atomic instance of a shortest paths network. In atomic instances there is an initial starting priority which is used for tie-breaking. If more than one player is trying to leave an edge and the capacity doesn't allow it, then the player with the better priority gets to exit first \cite{kochthesis, koch2011nash}.
\subsubsection{Atomic Shortest Paths Network Example}
\begin{figure}[H] 
\center{\includegraphics[width=0.75\linewidth]{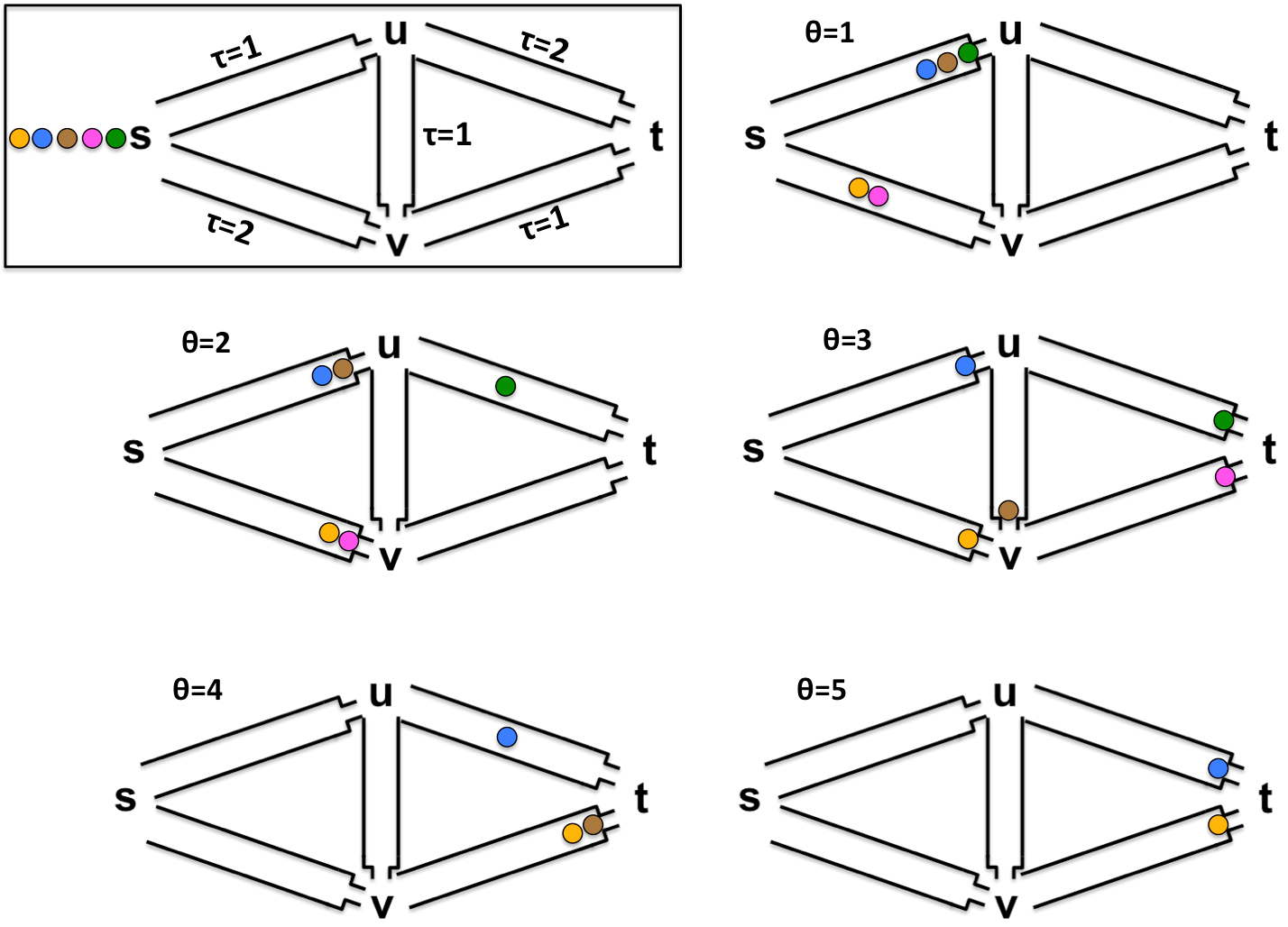}}
\caption{Atomic Shortest Paths Network}
\end{figure}
The top left square section of Figure 3 shows a dynamic network where each edge has capacity $1$. Free flow transit times are shown next to the edges, and five atomic players are shown in their initial starting order. Each player controls one unit of flow, thus only one player at a time can exit a given edge. We show the routing of the players at each point in time $\theta$ which results in a Nash Equilibrium. When referring to routes we will consider the $s\rightarrow u \rightarrow t$ path as the upper path, the $s\rightarrow v \rightarrow t$ path as the lower path, and the $s \rightarrow u \rightarrow v \rightarrow t$ path as the zig-zag path. The green player can choose any route, all result in an earliest arrival time of $\theta = 3$. For this example, the green player takes the upper route. The pink player will take the lower path to ensure an arrival time of $\theta = 3$. The brown player could take any of the three paths for an arrival time of $\theta=4$ because they must wait to exit either the $(s,u)$ edge or the $(s,v)$ edge because the pink and the green player are before them in the initial starting order. In this example the brown player takes the zig-zag path and arrives at $t$ at $\theta = 4$. The blue player can either take the upper or lower path; they cannot take the zig-zag path because then the orange player could take the lower path and enter the $(v,t)$ edge before them causing an arrival time of $\theta = 6$. So the blue player will take the upper edge for an arrival time of $\theta = 5$. Based on the blue player's choice of the upper edge, the orange player's quickest route is the lower edge and experiences an arrival time of $\theta = 5$.
\subsection{Random Queuing}
Popov and Tatarenko study a nonatomic random queuing model in \cite{RandomQueues}, where the network consists of a set of parallel paths that go from the source $s$ to the sink $t$. Similar to the shortest paths network, each edge has constant capacity, and all players start at the source at time $0$. Players can either traverse an edge, or stay waiting at their current vertex (if the edge capacity is reached) at each time step $\Delta t$. While this may seem different than the DQM at first glance, we can abstract the model by imagining each edge having a free flow traversal time of $1$, and instead of waiting at vertices players wait at the end of the edge leading into the vertex in which they would have been waiting at. When the capacity of an edge is reached and a waiting queue is formed, instead of players being allowed to exit in the normal FIFO manner players are chosen at random. The probability that a player is able to exit the edge they are at is given by the direct ratio of the capacity of the edge to the total amount of flow waiting to leave the edge. The social objective of the game is to minimize the sum of the arrival costs of all the players, where the cost of a player is their arrival time at the sink.
\subsubsection{Random Early Detection}
The Random Early Detection (RED) algorithm used for internet quality of service is an application of random queues in the deterministic queuing model. If a router is too busy and cannot deliver data efficiently, it may drop a packet. Dropping a packet serves as a signal to senders that the network is congested and they should either reduce the bandwidth they are using or try sending data along a different path. RED is used to drop packets before a queue becomes full. Each queue has a minimum threshold and a maximum threshold. No packets are dropped when the average queue size is less than the minimum threshold, and all packets are dropped or marked when the average queue size is larger than the maximum threshold. When the average queue size is between the minimum and maximum threshold, packets are marked/dropped probabilistically based on the average queue size \cite{InternetQOS}. This can be modeled by random queues in the deterministic queuing model because when an edge reaches capacity, flow particles are chosen at random to get to traverse to the next edge on their path. This means that there is a probability that a flow particle might not ever get chosen to exit an edge when capacity is reached, and would only be able to exit the edge once the amount of flow trying to exit that edge decreases below capacity, signifying a decrease in traffic. 
\subsection{Atomic Games with a Bottleneck Objective} 
In \cite{AtomicDQM}, Werth, Holzhauser, and Krumke examine the deterministic queuing model under atomic instances. In particular, they study atomic versions of the model in which players try to minimize their bottleneck value, and the overall objective is to minimize the bottleneck of the game. Since this is an atomic game, we need a tie-breaking scheme for when two players arrive at an edge at the same time and the capacity will not allow them both to exit at the same time. The tie-breaking scheme associated with this objective is a local tie-breaking scheme. In local tie-breaking, a priority is assigned to each edge entering a vertex. If more than one player arrives at the edge at the same time, then the players who are entering from an edge with a better priority get to exit first. The ordering between players who were on the same edge is carried over from the previous edges they traversed and each player is given a starting priority at the source vertex. \par
Another modification made in the study of atomic instances is allowing for multiple sources and sinks. The regular deterministic queuing model is a \textbf{single-commodity} instance since there is a single source and single sink. We can also consider instances in which there are \textbf{multiple commodities} and each player is associated with a commodity, as described in the selfish static routing model. While multi-commodity instances can be either nonatomic or atomic, \cite{AtomicDQM} was the only paper to consider multiple commodities in the deterministic queuing model. 
\subsubsection{Live Cargo Transportation} 
The European Union has many regulations surrounding the transportation of livestock. The regulations center around travel times and resting periods, and at each stop the livestock must be fed and inspected. This can be modeled through the deterministic queueing model, where an edge represents travel between two stops. The travel time between docks can be reflected in the free flow transit time, and the wait time reflects waiting in line to be serviced at a dock as there are many boats that cause congestion at a dock. When there is heavy traffic at the docks, there needs to be a way to break ties when boats arrive at the docks at the same time, and the current tie-breaking rules are similar to local tie-breaking. An indicator of livestock health is the time spent traveling between docks because it is a period of time that the livestock has to go without being serviced, and travel can be hard on the animals. Thus a good objective in this scenario is to minimize the bottleneck value, the longest time spent traversing an edge, which in this scenario would be the longest time the animals have to go without being serviced \cite{AtomicDQM}.
\subsubsection{All-Optical Networks} 
In all-optical networks, blocking occurs when more than one data packet arrives at a node at the same time. At each node, packets can be processed in a FIFO manner, however, sometimes there are local priorities based on the input ports of a node. This can be modeled with the local priority tie-breaking scheme. When blocking occurs, instead of a node dropping all but one packet, fibre delay lines are used to allow the packets to circle around until they can be forwarded to the next node. Packets deteriorate over time both on fibre delay lines and regular lines between nodes. Luckily, the packet contents can be regenerated when it is finally forwarded to the next node. Thus to ensure a good-quality routing, the objective would be to minimize the bottleneck of the game, i.e., the maximum time spent traveling between nodes and thus time in which the packet could deteriorate \cite{AtomicDQM}.
\section{Results} 
When studying selfish models, we not only want to measure how bad a Nash equilibrium is compared to optimal, but we want to know if a Nash equilibrium or optimal solution even exists, and if they can be computed. Even more interesting is how small modifications to the model or parameters affect these results on existence and computability. This section first examines results on the three objectives discussed in Section 2.2, and the general deterministic queuing model. It then investigates how the results change in the modified versions of the model.
\subsection{Optimal Flow Results}
Quickest flows on a single-commodity network are computed by doing a binary search on time. In a multi-commodity instance computation of a quickest flow is NP-Hard and it is unclear whether or not they always exist, but there does exist a 2-approximation algorithm for multi-commodity instances \cite{fleischer, skutellaIntroduction}. Earliest arrival flows always exist in the single-commodity instance and there is an exponential-time algorithm to compute them, there is also a polynomial-time approximation algorithm. While earliest arrival flows always exist in single-commodity instances, there are multi-commodity instances in which they do not exist \cite{skutellaIntroduction, schmidt}. The results are even worse for computing a narrowest flow in the atomic instance. In the multi-commodity case, for both weighted and unweighted players and in both cyclic and acyclic networks, computing a flow with minimum bottleneck value is NP-complete. This result holds true even for general single-commodity instances with weighted or unweighted players \cite{AtomicDQM}.
\subsection{Basic Deterministic Queuing Model and Shortest Paths Network} 
Koch and Skutella prove in \cite{koch2011nash} that every instance of a Nash flow in the deterministic queuing model can be seen as a concatenation of ``static thin flows \emph{with} resetting", and every instance of a Nash flow in the shortest paths network can be seen as a concatenation of ``static thin flows \emph{without} resetting".\footnote{For more information on static thin flows, see \cite{kochthesis}.} Based on existence results of static thin flows shown in \cite{kochthesis}, we can conclude that a Nash equilibrium always exists in instances of the deterministic queuing model and in instances of shortest paths networks. Since static thin flows \emph{without} resetting can be computed in polynomial time, this means that in the nonatomic version of shortest paths network a Nash equilibrium can be computed in polynomial time. This contrasts the results for an atomic version of the shortest paths network where there is an algorithm to compute Nash, and while the algorithm never terminates it can be stopped after exponential time and Nash can be derived from that with rounding. In shortest paths network not only does a Nash always exist, but that Nash is optimal (or asymptotically optimal in the atomic case). This means that the evacuation price of anarchy, total delay price of anarchy, and time price of anarchy are all 1 (asymptotically 1 in the atomic case). 
\subsubsection{Stackelberg Strategy} 
Currently, there are no universal bounds on the price of anarchy for the general deterministic queuing model, but Bhaskar, Fleisher, and Anshelevich generate upper bounds by enforcing a Stackelberg strategy \cite{Stackelberg}. A \textbf{Stackelberg strategy} adds a player to the game that acts as a network manager. This manager gets to route their flow ahead of all the other players, and thus reduces the capacities of the edges, and routes their flow with the same goal as the social objective of the game. In this model, the time price of anarchy is bounded from above by $e/(e-1)$ by having the network manager route an amount of flow that reduces the capacities to the same value on each edge as the static flow underlying the quickest flow. Such a Stackelberg strategy always exists and is polynomial-time computable. In the exact same manner, we can bound the total delay price of anarchy from above by $2e/(e-1)$ by having the network manager reduce the capacities on each edge to the value of that edge in the static flow underlying the earliest arrival flow. This strategy also always exists and is polynomial-time computable. Essentially, this strategy ensures that the static flow underlying the quickest (earliest arrival) flow saturates every edge of the graph. So if this is already the case of a given instance, then these bounds hold without needing to use this Stackelberg strategy.
\subsection{Random Queuing} 
In the random queuing version of the deterministic queuing model a Nash equilibrium always exists and can be computed by nonlinear convex programming. These Nash are not unique though, meaning that the sum of the arrival times of all the players can differ between different Nash equilibria. Recall that in the random queuing model, each player wants to minimize their arrival time at the sink and the social objective of the game is to minimize $C$, the sum of the arrival times of all the players. Given each player's routing path $P$, we can use a global planner to force each player to wait at the initial vertex for an amount of time equal to $.5\frac{f_P}{a_{min}(P)}\Delta t$ before starting their routing, where $f_P$ is the number of players using path $P$, and $a_{min}(P)$ is the minimum capacity on path $P$. Such a global planner ensures that \emph{one} of the Nash equilibrium resulting from this new routing has a cost within the range $[C^*, C^*+.5\Delta t]$, where $C^*$ is the minimal value of the social cost \cite{RandomQueues}. Note, that this is not a bound on the price of anarchy because the price of anarchy looks at the ratio between the worst-valued equilibrium and optimal, and this bound is only ensured for one resulting Nash, not all Nash equilibria.
\subsection{Atomic Games with a Bottleneck Objective} 
We can categorize different types of instances of atomic games with a bottleneck objective in terms of parameters such as whether players are weighted or not, whether the network is acyclic, and whether it is a multi-commodity or single-commodity instance. We applied theorems from \cite{AtomicDQM} to broader categories of instances, to obtain the results shown in Figure 4.
\begin{figure}[H] 
\begin{center}
	\begin{tabular}{| p{6cm} || p{8cm}|}
		\hline
		\textbf{Result} & \textbf{Type of Instances}\\
		\hline
		\hline
		There exist instances without a Nash Equilibrium. & Multi-commodity, cyclic or acyclic networks, with weighted or unweighted players.\\
		\hline
		Determining whether a given instance has a Nash equilibrium is NP-Hard in the strong sense. & Multi-commodity, cyclic or acyclic networks, with weighted or unweighted players. \\
		\hline
		Determining whether a given routing of all the players is a Nash is Co-NP-Complete. & Multi-commodity, cyclic or acyclic networks, with weighted or unweighted players. Single commodity, cyclic networks, with weighted or unweighted players. \\
		\hline
		Best Response computation is not guaranteed to find a Nash in every instance. & Single-commodity, cyclic or acyclic networks, with weighted or unweighted players. \\
		\hline
	\end{tabular}
\end{center}
\caption{Results for Specific Instances of Atomic Games with a Bottleneck Objective}
\end{figure} 
\noindent An interesting result from \cite{AtomicDQM} is that for all instances of atomic games with a bottleneck objective, the (bottleneck) price of anarchy is bounded from above by $k$, where $k$ is the number of players. A repercussion of players aiming to minimize their bottleneck value is that it results in an unbounded time, evacuation, and total delay price of anarchy. This is because when players are determining what path to take, paths with the same bottleneck value are indistinguishable to the player, however these paths could have very different total traversal times.  
\section{Conclusion} 
This paper presents the deterministic queuing model, its variations, applications, and results. We note that in some instances the model could be even further modified to better fit certain applications, and one would need to study how these new modifications affect the results. There are still gaps to be filled in the results of the deterministic queuing model and its modifications, along with other dynamic flow over time models. Since dynamic flows include time as a variable and parameters can vary with time, there is room to create even more models, and tailor those models to give sufficient solutions to even more real-world problems. 
\section{Acknowledgments}
I am grateful for my advisor Alexa Sharp for her support and wisdom, and to the Oberlin College Computer Science Department for giving me this research opportunity.
\printbibliography
\end{document}